\documentclass[hyper]{JHEP} 

\usepackage{epsfig}

\newcommand\fverb{\setbox\pippobox=\hbox\bgroup\verb}

\newcommand\fverbdo{\egroup\medskip\noindent%

            \fbox{\unhbox\pippobox}\ }

\newcommand\fverbit{\egroup\item[\fbox{\unhbox\pippobox}]}

\newbox\pippobox

\title{D1-brane with Overcritical Electric Field in AdS$_3$ and  S-brane}
\author{J. Kluso\v{n}\\
Dipartimento di Fisica,\\
Universita' di Roma \& I.N.F.N. Sezione di Roma 2, ``Tor Vergata'' \\
Via della Ricerca Scientifica, 1 00133  Roma,   ITALY\\
E-mail: \email{Josef.Kluson@roma2.infn.it}}
\author{Rashmi R. Nayak \\
Dipartimento di Fisica (DIFI),\\
Universita' di Genova \& I.N.F.N. Sezione di  ``Genova''\\
Via Dodecaneso 33, 16146 Genova, ITALY\\
E-mail: \email{Rashmi.Nayak@ge.infn.it}}
\author{Kamal L. Panigrahi\\
Physics Department\\
Indian Institute of Technology, Guwahati, INDIA, 781 039 \\
\& \\
Dipartimento di Fisica, Universita' di ``Genova'', Genova, ITALY \\
E-mail: \email{panigrahi@iitg.ernet.in,
Kamal.Panigrahi@ge.infn.it}} \preprint{GEF-TH-04/2006\\
ROM2F/2006/06 \\ \hepth{0602211}} \abstract{We study aspects of
Dirichlet S-branes, which are defined as Dirichlet boundary
condition on a time like embedding of open strings, in general
backgrounds. By applying $T$-duality along an isometry of the
unphysical dS$_2$-branes in NS-NS supported AdS$_3$-background, we
find S0-brane. We also study the time dependent tachyon
condensation on the unstable Dp-brane and interpret the singular
solutions as lower dimensional S(p-1)-brane that couples to real
Ramond-Ramond fields while to imaginary NS-NS modes.}

\def\bA{\mathbf{A}}

\def\bAi{\left(\mathbf{A}^{-1}\right)}

\def\mL{\mathcal{L}}

\def\mF{\mathcal{F}}
\def\mat{\mathbf{a}}

\def\mtF{\tilde{\mathcal{F}}}
\def\mtC{\tilde{\mathcal{C}}}
\def\mati{(\mathbf{a}^{-1})}
\begin{document}
\section{Introduction}\label{first}

Space-like branes or S-branes \cite{Gutperle:2002ai} are
fascinating objects in string theory. They are defined as a kind
of topological defects localized on a space-like hypersurface, and
hence can only exist for `moment' in time. The rolling of the open
string tachyon on an unstable D-brane (for a review see
\cite{Sen:2004nf} and references therein) namely the decaying
branes to the closed string vacuum in some cases can be considered
as an array of Dirichlet S-branes in imaginary time
\cite{Gaiotto:2003rm}. In general S-branes can also be viewed as
the time-dependent homogeneous solutions in string theory or in
supergravity, localized in a given instant of time. They have been
very useful in understanding cosmological applications of string
theory.

Dirichlet S-branes are also obtained by imposing a Dirichlet
condition on the time like coordinate of the open strings
\cite{Durin:2005ts}. Under T-duality along a transverse spatial
direction the S-branes are shown to be T-dual to the D-branes with
overcritical electric field. It was further observed that unlike
the D-branes, in the first quantization of the open string between
a pair of S-branes, there are only a finite number of physical
states that increases when they gets separated with time. In
general S-brane solutions in the type-II string theory can be
obtained by analytically continuing the usual D-brane boundary
states, but one has to keep in mind that it radiates the
Ramond-Ramond field with wrong reality property. In other words
one can have a S-brane with real R-R charge, but then it won't be
a solution of type-II theories rather its existence can be
predicted in II* theory. Further it was shown that the generic
S-brane configurations should decay into a bunch of D-branes (or
brane-anti-brane pairs).

D-branes in the Anti-de Sitter backgrounds have been studied by
various authors in the past by using various techniques, see for
example (\cite{Bachas:2000fr}-\cite{Hikida:2005vd}). String theory
on the SL(2,R) and its discrete orbifolds have shed new light in
the conjectured AdS/CFT duality. The corresponding target space
geometry is AdS$_3$ supported by NS-NS three-form flux. D-branes
in this background has been considered in the past. In
\cite{Bachas:2000fr}, it was shown that the $dS_2$-branes in
AdS$_3$ are unphysical due to the presence of overcritical
electric field. So it is tempting to examine the behavior of these
unphysical D-branes in the T-dual background in the light of
\cite{Durin:2005ts}. We address this question in this paper. In
doing so, what we achieve is the following. First of all we are
able to find a physical interpretation of the unphysical solutions
with imaginary electric flux etc as in the T-dual picture the
S0-brane that arises from the time dependent tachyon condensation
on unstable D1-brane. Second, the previously found unphysical
solutions correspond, in fact, to perfect and acceptable solutions
in string theory (even if the initial configurations of tachyon
that corresponds to S-brane have to be fine tuned) since they
arise from the open string tachyon condensation.

The rest of the paper is organized as follows. In section-2 we try
to spell out some properties of the Anti-de Sitter D-branes and
show the unphysicalness of the $dS_2$-branes in AdS$_3$. In
section-3, we apply $T$-duality along one of the symmetry
directions, and interpret the solution that can be seen as
D0-brane moving along that particular direction. We find out the
equation of motion for the dynamical variable for the later
comparison with the S0-branes. In section-4, we study the time
dependent tachyon condensation on the unstable D-brane and found
out the signature of the S$(p-1)$ branes. In fact, we found that
the dynamics of the kink is governed by the equations of motion
that arise from the S-brane effective action in given background
\footnote{S-brane action was also studied in
\cite{Hashimoto:2002sk,Dorn:2005vg,Bhattacharya:2003iw,
Hashimoto:2003qx}.}. We further analyze the properties of
energy-momentum tensor derived from such DBI action. The main
result of this analysis is the fact that the singular time
dependent tachyon condensation on an unstable Dp-brane leads to
the emergence of the object whose equations of motion arises from
the action that can be interpreted as S(p-1)-brane with imaginary
tension (in other words, it couples to imaginary NS-NS modes) and
with real charge with respect to Ramond-Ramond fields. This is
equivalent to the analysis performed in \cite{Durin:2005ts}  where
this kind of Sp-branes was named as $S^-p$-brane \footnote{It is
necessary to mention one important  subtlety considering our
results and the work \cite{Durin:2005ts}. It was argued there that
the $S^-p$-brane should contain open string tachyon in its world
volume theory. Unfortunately using effective field theory
description performed below we are not be able to find the
evidence for the existence of this tachyon. It is of course
possible that more general ansatz for fluctuations around the time
dependent tachyon solution of the non-BPS  Dp-brane world volume
theory will contain in its spectrum a tachyonic mode. We hope to
return to this problem in future.}. Then we apply this general
procedure in section-5 to find out the S0-brane equations of
motion that resembles with that of the D-brane in the dual
background. Finally in section-6, we present our conclusions.

\section{D-branes with overcritical electric fields and emerging S-branes}
In this section we study the properties of the D-branes with
overcritical electric fields. Let us begin with the $AdS_3\times
S^3$ metric in global coordinates:
\begin{equation}\label{bachasads}
ds^2=L^2[-\cosh^2\rho dt^2+
d\rho^2+\sinh^2\rho d\theta_1^2]+
L^2[d\theta^2+\cosh^2\theta d\tilde{\psi}^2+
\sin^2\theta d\theta_2^2]
\end{equation}
supported by the Neveu-Schwarz three-form field
\begin{equation}
H=dB=L^2\sinh(2\rho)d\rho\wedge d\theta_1
\wedge dt \ ,
B=L^2\sinh^2\rho d\theta_1\wedge dt \ .
\end{equation}
Let us consider the D1-brane in the above background with the DBI
action
\begin{equation}\label{actBPS}
S=-\tau_1\int d^2\xi
e^{-\Phi}
\sqrt{-\det\bA} \ ,
\end{equation}
with
\begin{equation}
\bA_{\mu\nu}=g_{MN}\partial_\mu X^M
\partial_\nu X^N+b_{MN}
\partial_\mu X^M
\partial_\nu X^N+
(2\pi\alpha')(\partial_\mu A_\nu-\partial_\nu A_\mu) \ ,
\end{equation}
where
Dp-brane tension is equal to
\begin{equation}
\tau_p=\frac{1}{(2\pi)^{ \frac{(p-1)}{2}}(2\pi\alpha')^{\frac{
p+1}{2}}} \  ,
\end{equation}
$X^M \ , M=0,\dots,9$ label the position of D1-brane, $g_{MN} \ ,
b_{MN}$ are background metric and NS-NS two form field
respectively and $A_\mu \ , \mu=0,1$ is worldvolume gauge field.
Now the equations of motion for $X^K$
derived from the  action
(\ref{actBPS})
 take the
form {\small
\begin{eqnarray}\label{eqXKg}
&&\partial_K[\tau_p e^{-\Phi}]\sqrt{-\det\bA}+
\frac{\tau_pe^{-\Phi}}{2} \left[\partial_Kg_{MN}+\partial_Kb_{MN}
\right]\partial_\mu X^M
\partial_\nu X^N\bAi^{\nu\mu}
\sqrt{-\det\bA}-\nonumber \\
&&-\partial_\mu \left[\tau_p e^{-\Phi} \left\{g_{KM}\partial_\nu
X^M\bAi^{\nu\mu}_S + b_{KM}\partial_\nu X^M
\bAi^{\nu\mu}_A\right\} \sqrt{-\det\bA}\right]=0 \ ,
\nonumber \\
\end{eqnarray}}
while the equation of motion for
 the gauge field $A_\nu$ takes the form
\begin{equation}
\partial_\mu\left[
\tau_p e^{-\Phi}
(2\pi\alpha')\bAi^{\nu\mu}_A
\sqrt{-\det\bA}\right]=0 \ ,
\end{equation}
where the symmetric and anti-symmetric part of the
$\bAi^{\nu\mu}$, respectively, are given by
\begin{equation}
\bAi^{\nu\mu}_S=
\frac{1}{2}
\left(\bAi^{\nu\mu}+\bAi^{\mu\nu}\right) \ ,
\bAi^{\nu\mu}_A=
\frac{1}{2}
\left(\bAi^{\nu\mu}-\bAi^{\mu\nu}\right) \ .
\end{equation}
Let us now consider the D1-brane that wraps $\theta_1$ direction
and study its dynamics when all the worldvolume modes depend on
time only. More precisely, we fix the gauge as
\begin{equation}\label{fa}
\theta_1=\xi^1 \ ,
\xi^0=t=X^0 \
\end{equation}
and also take $A_0=0$. Let us also presume that $\rho=\rho(t)$.
Then the matrix $\bA$ is equal to
\begin{eqnarray}
&&\bA_{00}
=-L^2\cosh^2\rho+L^2\dot{\rho}^2 \ ,
\nonumber \\
&&\bA_{01}
=-L^2\sinh^2\rho+
(2\pi\alpha')\dot{A}_{\theta_1} \ , \nonumber \\
&&\bA_{10}
=L^2\sinh^2\rho-(2\pi\alpha')\dot{A}_{\theta_1}  \ ,
\nonumber \\
&&\bA_{11}
=L^2\sinh^2\rho \ ,
\nonumber \\
\end{eqnarray}
where $\dot{f}=\frac{df}{dt}$.
Consequently we get
\begin{equation}
\det\bA=
\det \tilde{g}+\mF_{t\theta_1}^2 \ ,
\end{equation}
where
\begin{equation}
\det \tilde{g}=-L^4\cosh^2\rho\sinh^2\rho+
L^4\sinh^2\rho\dot{\rho}^2 \ , \mathcal{F}_{t\theta_1}=-
L^2\sinh^2\rho+(2\pi\alpha')\dot{A}_{\theta_1} \
\end{equation}
and also
\begin{equation}
\bAi=\frac{1}{\det\bA}
\left(\begin{array}{cc}
\tilde{g}_{\theta_1\theta_1} & -\mF_{t\theta_1} \\
\mF_{t\theta_1} & \tilde{g}_{tt} \\
\end{array}\right)
\end{equation}
Now the equation of motion for $A$ gives
\begin{equation}\label{mF}
\frac{(2\pi\alpha')\tau_1\mF_{t\theta_1}}
{g_s\sqrt{-\det\tilde{g}-
\mF_{t\theta_1}^2}}=-q
\Rightarrow
\mF_{t\theta_1}^2=-\frac{g_s^2q^2\det\tilde{g}}
{g_s^2q^2+(2\pi\alpha')^2\tau_1^2} \ .
\end{equation}
We must also check that the equation of motion (\ref{eqXKg}) are
obeyed for the ansatz (\ref{fa}). For $K=\theta_1$ the equation
(\ref{eqXKg}) is trivially satisfied since now all the modes do
not depend on $\theta_1$. On the other hand the equation of motion
for $X^0$ gives
\begin{equation}
\partial_0
\left[\frac{\tau_1}{g_s}g_{tt}\bAi^{00}_S
\sqrt{-\det\bA}\right]+
\partial_0\left[\frac{\tau_1}{g_s}
b_{t\theta_1}\bAi^{\theta_1 t}_A
\sqrt{-\det\bA}\right]=0 \
\end{equation}
and hence we obtain the conserved quantity
\begin{eqnarray}\label{Ed1}
\frac{E}{2\pi}
=\frac{- g_{tt}g_{\theta_1\theta_1}
\sqrt{q^2+(2\pi\alpha')^2\tau_1^2g_s^{-2}} -q
b_{t\theta_1}\sqrt{-\det\tilde{g}}}
{(2\pi\alpha')\sqrt{-\det\tilde{g}}}
\nonumber \\
\end{eqnarray}
using (\ref{mF}). Some comments regarding the definition of the
conserved quantity $E$ is in order now. Here $E$ means the
conserved energy of the D1-brane that arises by simply integrating
over $\theta_1$ direction which implies (for homogeneous
worldvolume fields) that $E$ is proportional to $2\pi$. We have
further included the factor $e^{-\Phi}=e^{-\Phi_0}=\frac{1}{g_s}$,
where $g_s$ is the string coupling constant. This factor is
important for the later comparison with the D0-brane equations of
motion which we derive in the next section.

Let us try to evaluate the energy on the solution
\cite{Bachas:2000fr}
\begin{equation}
\cosh \rho\cos t=C \ , C>0 \ .
\end{equation}
Firstly, we have
\begin{equation}
-\det\tilde{g}= \frac{L^4C^2}{\cos^4 t} (C^2-1) \ .
\end{equation}
Then the expression for energy is
\begin{eqnarray}
\frac{E}{2\pi}=
L^2\sinh^2\rho\left(\frac{C\sqrt{q^2(2\pi\alpha')^{-2}
+\tau_1^2g_s^{-2}}+
q\sqrt{C^2-1}}
{\sqrt{C^2-1}}\right) \ .
\end{eqnarray}
As we have determined above $E$ has to be conserved, but on the
other hand we see that it depends explicitly on $\sinh\rho$. So
the only possibility for it to be conserved is that it has to
vanish. This occurs when
\begin{equation}
C\sqrt{q^2(2\pi\alpha')^{-2}+\tau_1^2g_s^{-2}}+
\frac{q}{(2\pi\alpha')}\sqrt{C^2-1}=0
\end{equation}
and this implies
\begin{equation}
q^2=-C^2(2\pi\alpha')^2\frac{\tau_1^2}{g_s^2} \ .
\end{equation}
So we obtain the well known result that the dS$_2$-brane
corresponds to the unphysical situation when the electric flux on
its worldvolume is purely imaginary.

Let us now return to the equation of conserved energy and try to
solve it explicitly. Using the conserved
energy given in (\ref{Ed1}) we get
\begin{eqnarray}\label{dotrhod1}
\dot{\rho}^2=- \frac{\cosh^4\rho\sinh^2\rho (\frac{L^4\tau_0^2}{
g_s^{2}\alpha'}+\frac{q^2L^4}{\alpha'^{2}})}
{(E-\frac{L^2}{\alpha'}q\sinh^2\rho)^2}+
\cosh^2\rho
\end{eqnarray}
using
\begin{equation}
\tau_1^2=\frac{1}{4\pi^2\alpha'}\tau_0^2 \ .
\end{equation}

The differential equation above can be solved explicitly,
however it leads to the very complicated result, which we don't
wish to present here. We will only briefly discuss
the properties of given solution when we
presume
that  $E,q$ are real and also $\dot{\rho}^2>0$.
Then the equation (\ref{dotrhod1}) implies
following bound for $\rho$
\begin{eqnarray}
-\frac{L^4\cosh^4\rho\sinh^2\rho
(\tau_1^2g_s^{-2}+(2\pi\alpha')^{-2}q^2)}
{(\frac{E}{2\pi}-q(2\pi\alpha')^{-1}L^2\sinh^2\rho)^2}+
\cosh^2\rho>0 \ .
\end{eqnarray}
Solving this inequality leads to the condition
\begin{equation}
 \sinh^2\rho\in (0,\sinh^2\rho_+)  \ ,
\end{equation}
where $\sinh^2\rho_+$ is a root of the quadratic equation given
above. In other words, for real $E$ and $q$, we obtain motion in
the finite interval and D1-brane cannot reach the boundary of
$AdS_3$.

Instead of studying the properties of the classical trajectory of
D1-brane in more detail we rather turn our attention to the
possibility of explaining these unphysical solutions with
imaginary electric flux in the $T$-dual set up.

\section{T-dual Background}\label{third}
On the other hand it was argued recently that such a configuration
could be related to $T$-dual situation where it could correspond
to Dirichlet S-brane. To make this statement more clear and
precise, let us apply $T$-duality along $\theta_1$ direction. More
precisely, the action of $T$ -duality along the symmetry direction
$\theta_1$ maps the string frame metric to string frame metric
\cite{Giveon:1994fu}
\begin{eqnarray}\label{Tdualback}
&&d^2\tilde{s}=\alpha' [g_{\mu\nu}-\frac{1}{g_{\theta_1\theta_1}}
(g_{\mu \theta_1}g_{\theta_1\nu}- B_{\mu \theta_1}B_{\nu \theta_1})]dx^\mu
dx^\nu+ 2\frac{1}{g_{\theta_1\theta_1}}B_{\theta_1\mu}d\theta_1 dx^\nu+
\frac{1}{g_{\theta_1\theta_1}}d\theta_1^2 \ , \nonumber \\
&&\tilde{B}=\frac{\alpha'}{2}dx^\mu \wedge dx^\nu
[B_{\mu\nu}-\frac{1}{g_{\theta_1\theta_1}} (g_{\mu
\theta_1}B_{\theta_1\nu}+ B_{\mu \theta_1}g_{\theta_1\nu})]+
\frac{\alpha'}{g_{\theta_1\theta_1}}
g_{\theta_1\mu}d\theta_1\wedge dx^\mu \ , \nonumber \\
&&\tilde{\phi}=\phi-\frac{1}{2}
\log g_{\theta_1\theta_1} \ . \nonumber \\
\end{eqnarray}
where we have included in the original components of the metric
$g_{\mu\nu}$ and the anti-symmetric tensor $B_{\mu\nu}$, the
dimensionless factor $\frac{L^2}{\alpha'}$.

Now we are ready to perform the T-duality along $\theta_1$
directions. Recall that in our convention $\theta_1$ is
dimensionless and periodic with period $2\pi$. In T-dual
background we rename $\theta_1$ as $z$ that is still periodic with
period $2\pi$. Finally we write T-dual components of the metric
$g$ and the anti-symmetric $B$ with the factor $\alpha'$. Then
the metric components of the dual background take the form (We
denote the dual variable to $\theta_1$ as $z$)
\begin{eqnarray}\label{Tdualads}
\tilde{g}_{zz}= \frac{\alpha'^2}{L^2\sinh^2\rho} \ ,
\tilde{g}_{tt}=-L^2 \ ,\tilde{g}_{tz}=\tilde{g}_{zt}=\alpha' \ ,
\tilde{g}_{\rho\rho}=L^2 \nonumber \\
\end{eqnarray}
while the other components of the metric remain unchanged. We also
get new components of the anti-symmetric $B$ field
\begin{equation}
\tilde{B}_{zt}=-L^2 \ .
\end{equation}
Finally, we also obtain nonzero value of the dilaton in the form
\begin{equation}
\tilde{\phi}=\phi_0-\frac{1}{2}\ln g_{\theta_1\theta_1}=
\phi_0-\ln \frac{L}{\sqrt{\alpha'}}\sinh\rho \ .
\end{equation}
Under $T$-duality the D1-brane that wraps the circle is mapped to
the D0-brane that moves around this circle. Recall that dynamics
of the D0-brane is governed by the action
\begin{equation}\label{Sact}
S=-\tau_0\int d\tau e^{-\Phi}
\sqrt{-\bA} \ ,
\bA=g_{MN}\dot{X}^M\dot{X}^N \ ,
\end{equation}
where in the following we omit the tilde on $g$. The equations of
motion for $X^M$ that follow from the action (\ref{Sact}) take the form
\begin{equation}
\partial_K \left[e^{-\Phi}\right]
\sqrt{-\bA}
-\frac{1}{2}e^{-\Phi}\partial_K g_{MN}
\dot{X}^M\dot{X}^N\frac{1}{\sqrt{-\bA}}
+\frac{d}{d\tau}
\left[e^{-\Phi}\frac{
g_{KM}\dot{X}^M}{
\sqrt{-\bA}}\right]
=0 \ .
\end{equation}
Now we fix the gauge that the worldvolume parameter $\tau$ is
equal to $t\equiv X^0$. Then $\bA$ is equal to
\begin{equation}
\bA=g_{tt}+g_{\rho\rho}
\dot{\rho}^2+2g_{tz}\dot{Z}
+g_{zz}\dot{Z}^2 \
\end{equation}
and also the  equation of motion for $X^0=\tau$
takes the form
\begin{equation}
\frac{d}{d\tau}
\left[\frac{e^{-\Phi}(g_{00}
+g_{tz}\dot{Z})}{\sqrt{-\bA}}\right]=0
\end{equation}
that implies that
the quantity in the bracket is conserved. As usual it is
useful to make use of the Hamiltonian formalism after fixing the
 gauge.
To do this we  observe that the Lagrangian has the form
\begin{equation}
\mL=-\sqrt{V-\sum_i (f_i
(\partial_0{\Phi}^i)^2
+B_i\partial_0\Phi^i)}\equiv -\triangle \ ,
\end{equation}
where $V$ contain scalar potential for various fields $\Phi^i$.
The conjugate momentum $P_i$ to $ \Phi_i$ takes the form
\begin{equation}
P_i=\frac{\delta \mL}{\delta \partial_0\Phi^i}=
\frac{2f_i\partial_0\Phi^i+B_i}{2\triangle}
\ , \partial_0\Phi^i=\frac{1}{2f_i}
\left(2P_i\triangle-B_i\right)
\end{equation}
so that the Hamiltonian  is equal to
\begin{eqnarray}
H=\sum_iP_i\partial_0\Phi^i-\mL=
\frac{2V+\sum\frac{B_i^2}{2f_i}}{2\triangle}
-\sum_i
\frac{B_i}{2f_i} P_i=\nonumber \\
=\sqrt{\left(V+\sum_i\frac{B_i^2}{4f_i}\right)
\left(1+\sum_i\frac{P^2_i}{f_i}\right)} -
\sum_i\frac{B_iP_i}{2f_i} \ ,
\nonumber \\
\end{eqnarray}
where on the second line we have expressed the Hamiltonian as a
function of canonical variables $\Phi^i,P_i$.
Returning to the action  (\ref{Sact})
we obtain
\begin{eqnarray}
V=-e^{-2\Phi}\tau_0^2g_{tt} \ ,
f_z=e^{-2\Phi}\tau_0^2
g_{zz} \ , f_{\rho}=
e^{-2\Phi}\tau_0^2g_{\rho\rho}
 \ ,
B_{z}=2e^{-2\Phi}\tau_0^2g_{zt}
\nonumber \\
\end{eqnarray}
and hence the Hamiltonian is equal to
\begin{equation}\label{Hd0}
H=
\frac{1}{\sqrt{g_{zz}}}
\sqrt{(-g_{tt}g_{zz}+g_{tz}^2)
\left(e^{-2\Phi}\tau_0^2+\frac{1}{g_{zz}}P^2_z+
\frac{1}{g_{\rho\rho}}P^2_\rho\right)}
-\frac{g_{zt}}{g_{zz}}P_z \ .
\end{equation}
Firstly, since the Hamiltonian does not explicitly depend on $Z$
it implies that $P_z$ is constant of motion
\begin{equation}
\dot{P}_z=-\frac{\delta H}{\delta Z}=0 \ .
\end{equation}
On the other hand the equation of motion for $\rho$ is
\begin{equation}
\dot{\rho}=
\frac{\delta H}{\delta P_\rho}=
\frac{1}{g_{zz}g_{\rho\rho}}
(-g_{tt}g_{zz}+g_{tz}^2)\frac{P_\rho}
{E+\frac{g_{zt}}{g_{zz}}P_z} \ .
\end{equation}
As usual we simplify this equation using the fact that the
Hamiltonian is conserved and equal to energy $E$. Then
we express from (\ref{Hd0}) $P_\rho$ as
\begin{eqnarray}
P^2_\rho=
\frac{1}{1+\sinh^2\rho}
(E+\frac{L^2}{\alpha'}P_z\sinh^2\rho)^2
-\left(\frac{L^4\tau_0^2}{\alpha'g_s^2}
\sinh^2\rho+
\frac{L^4}{\alpha'^2}P_z^2\sinh^2\rho\right)
\nonumber \\
\end{eqnarray}
using the explicit metric components given above and also the fact
that $e^{-2\Phi}=\frac{1}{g_s^2} \sinh^2\rho$.
Then
we obtain
\begin{eqnarray}\label{dotrhod0}
\dot{\rho}^2
=-\frac{\cosh^4\rho\sinh^2\rho}
{(E+\frac{L^2}{\alpha'}P_z\sinh^2\rho)^2}
(\frac{L^4\tau_0^2}{\alpha'g_s^2}+ \frac{L^4}{\alpha'^2}P_z^2)
+\cosh^2\rho \ .
\nonumber \\
\end{eqnarray}
Now the equation (\ref{dotrhod0}) describes the dynamics of
D0-brane in dual background. As we expect this equation is the
same as the equation that determines the dynamics of D1-brane in
the original background. In fact, we see that this has the same
form as the equation (\ref{dotrhod1}) if we identify
\begin{equation}\label{Pzq}
P^2_z=q^2 \ .
\end{equation}
Naively we can say that this 
is the correct quantization condition for the motion of
a test D0-brane along a compact direction of periodicity $2\pi$. 
Of course there is an important issue that the momentum
is imaginary and hence the wave function of D0-brane
 is not periodic in $z$ variable\footnote{In any case, if one computes 
the squared norm of the tangent vector to the D0-brane trajectory,
this gives $\left(-\frac{L^2\cosh^4\rho}{P^2_z g^2_s \sinh^2 \rho}\right)$. 
Therefore if $P_z$ is imaginary this corresponds to supernuminal signature.}.
We should rather claim that the momentum $P_z$ is
conserved with the value given in (\ref{Pzq}).

We can
also see that the energy is the same in both cases. Then it
immediately follows that the
 classical trajectory $\cosh\rho\cos t=C$ which
corresponds to imaginary $P_z$  is again unphysical
while $E$ is equal to zero.
Note also that 
the equation of motion for $z$ takes
the form
\begin{equation}\label{dotZ}
\dot{Z}=
\frac{\delta H}{\delta P_z}=
\frac{1}{g_{zz}^{2}}(-g_{tt}g_{zz}+
g_{tz}^2)\frac{P_z}{E+\frac{g_{zt}}{g_{zz}}P_z}-\frac{g_{zt}}{g_{zz}}
\end{equation}
that for $E=0 \ , P_z=-q$, reduces to
\begin{equation}\label{dotZE0}
\dot{Z}=-\frac{g_{tt}}{g_{zz}}=\frac{L^2}{\alpha'} \ .
\end{equation} 
We see that the velocity $v_z$ is constant
and it does not depend on the value of the 
charge $q$. This result is a consequence
of the fact that the energy $E$ is zero 
for the trajectory $\cosh\rho\cos t=C$
as can be seen easily from the form of the
equation (\ref{dotZ}). On the other hand
it is also clear that when
$E\neq 0$ the motion along $z$ direction
will depend on $P_z$.

In summary, in the $T$-dual picture in case of ordinary D0-brane
we once again obtain a situation which is unphysical. Then, following
\cite{Durin:2005ts} we can expect that in $T$-dual background the
object, that is obtained in the dual picture of the corresponding
$D1$-brane will be a $S0$-brane. To see this explicitly we perform
in the next section the analysis of the time-dependent tachyon
condensation on unstable D$p$-brane in general background. We
argue that there exists a singular time dependent tachyon solution
which leads to the emergence of $S(p-1)$-brane that has imaginary
charge with respect to NS-NS fields however has real charge with
respect to the Ramond-Ramond (RR) fields.

\section{S(p-1)-brane in General Background}
This section is devoted to the study of the singular time
dependent tachyon condensation on the world volume of non-BPS
Dp-brane that leads to the emergence of S(p-1)-brane.

Once again, we begin with the Dirac-Born-Infeld like tachyon
effective action in general background
\cite{Sen:1999md,Garousi:2000tr,Bergshoeff:2000dq,Kluson:2000iy}
\footnote{We will work in this section in units
$(2\pi\alpha')=1$.}
\begin{eqnarray}\label{acg}
&&S=-\int d^{p+1}\xi e^{-\Phi}V(T)\sqrt{-\det \bA} \ ,
\nonumber \\
&&\bA_{\mu\nu}=g_{MN}
\partial_\mu X^M\partial_\nu X^N+
b_{MN}\partial_\mu X^M\partial_\nu X^N+
 F_{\mu\nu}+
\partial_\mu T\partial_\nu T \ ,
\mu \ , \nu=0,\dots, p \ ,
\nonumber \\
&&F_{\mu\nu}=\partial_\mu A_\nu -\partial_\nu A_\mu \ ,
\nonumber \\
\end{eqnarray}
where $A_\mu \ , \mu,\nu=0,\dots,p$ and $ X^{M,N} \ , M,
N=0,\dots,9$ are gauge and the transverse scalar fields on the
worldvolume of the non-BPS Dp-brane and $T$ is the tachyon field.
$V(T)$ is the tachyon potential that is symmetric under
$T\rightarrow -T$ has maximum at $T=0$ equal to the tension of a
non-BPS Dp-brane $\tau_p$ and has its minimum at $T=\pm \infty$
where it vanishes.

We must also stress that there exists a Wess-Zumino term for
non-BPS Dp-brane that expresses the coupling of this Dp-brane to
the Ramond-Ramond fields. \cite{Kennedy:1999nn,
Billo:1999tv,Kraus:2000nj,Takayanagi:2000rz,Okuyama:2003wm} that
is expected to have the form
\begin{equation}\label{SWZ}
S_{WZ}=\int_{\Sigma}
V(T)dT\wedge Ce^{F+B} \ ,
\end{equation}
where $\Sigma$ denotes the worldvolume of
non-BPS Dp-brane and $C$ collects all RR $n$-form
gauge  potentials (pulled back to brane worldvolume).

In what follows we closely follow the analysis
performed in  \cite{Kluson:2005fj}.
As usual    we start to solve the equations of motion for $T,X^M$ and
$A_\mu$. The equation of motion for tachyon takes the form
\begin{equation}\label{eqT}
-e^{-\Phi}V'(T)\sqrt{-\det\bA}+
\partial_\mu
\left[e^{-\Phi}\partial_\nu T
\bAi^{\nu\mu}_S\sqrt{-\det\bA}\right]+J_T=0 \ ,
\end{equation}
where $J_T=\frac{\delta }{\delta T}S_{WZ}$.
For scalar modes we obtain
\begin{eqnarray}\label{eqX}
&&-\frac{\delta e^{-\Phi}}
{\delta X^K}V\sqrt{-\det\bA}-\nonumber \\
&&-\frac{e^{-\Phi}}{2} V\left(\frac{\delta g_{MN}}{\delta X^K}
\partial_\mu X^M\partial_\nu X^N+
\frac{\delta b_{MN}}{\delta X^K}
\partial_\mu X^M\partial_\nu X^N\right)
\bAi^{\nu\mu}\sqrt{-\det\bA}+
\nonumber \\
&&+
\partial_\mu\left[e^{-\Phi}V
g_{KM}\partial_\nu X^M
\bAi^{\nu\mu}_S
\sqrt{-\det\bA}\right]
+\nonumber \\
&&
\partial_\mu \left[
e^{-\Phi}Vb_{KM}\partial_\nu X^M
\bAi^{\nu\mu}_A
\sqrt{-\det\bA}\right]+J_K=0  \ ,
\nonumber \\
\end{eqnarray}
where $J_K=\frac{\delta }{\delta X^K}S_{WZ}$.
Finally, the equations of motion for $A_\mu$ are given by
\begin{equation}\label{eqA}
\partial_\nu
\left[e^{-\Phi}V
\bAi^{\mu \nu}_A\sqrt{-\det\bA}\right]+J^\mu=0 \ ,
\end{equation}
where $J^\mu=\frac{\delta}{\delta A_\mu}S_{WZ}$. Now we try to
find the solution  of the  equations of motion (\ref{eqT}),
(\ref{eqX}) and (\ref{eqA}) that can be interpreted as a lower
dimensional S(p-1)-brane. More precisely, we can show that the
dynamics of the kink is governed by the equations of motion that
arise from the action for S(p-1)-brane in general background
\begin{eqnarray}\label{Sactg}
&&S=S_{DBI}^S+S_{WZ}^S  \ , \nonumber \\
&&S_{DBI}^S=-T_{S(p-1)}\int d^p\xi e^{-\Phi}\sqrt{\det\mat} \ ,
\nonumber \\
&&S_{WZ}^S=\mu_{S(p-1)} \sum_{n\geq 0} \frac{1}{n!(2!)^n(2p-2n)!}\int
d^p\xi \epsilon^{\alpha_1\dots \alpha_p} (\mtF)^n_{\alpha_1\dots
\alpha_{2n}} \mtC_{\alpha_{2n+1}\dots \alpha_p}
 \ , \nonumber \\
\end{eqnarray}
where
\begin{eqnarray}
&&\mat_{\alpha\beta}= (g_{MN}+b_{MN})\partial_\alpha
X^M\partial_\beta X^N+
F_{\alpha\beta} \ , \nonumber \\
&&\mtF_{\alpha\beta}=F_{\alpha\beta}+ b_{MN}\partial_\alpha
X^M\partial_\beta
X^N \  \ , \nonumber \\
&&\mtC_{\alpha_{2n+1}\dots \alpha_p}= C_{M_{2n+1}\dots
M_p}\partial_{\alpha_{2n+1}}
X^{M_{2n+1}}\dots\partial_{\alpha_p}X^{M_p} \
\nonumber \\
\end{eqnarray}
and $\xi^\alpha,\alpha=1,\dots,p$.
Finally, $T_{S(p-1)}$ is S(p-1)-brane
tension and $\mu_{S(p-1)}$ is the
charge of S(p-1)-brane with respect to
Ramond-Ramond fields. These quantities
will be determined during the calculations.

In other words we will show that the  modes given in (\ref{ansA})
that propagate on the worldvolume of the kink obey the  equations
of motion derived from (\ref{Sactg}) that have the form
\begin{eqnarray}\label{eqXbps}
-T_{S(p-1)}
\frac{\delta e^{-\Phi}}{\delta X^K}
\sqrt{\det\mat}-\nonumber \\
-T_{S(p-1)}\frac{e^{-\Phi}}{2}
\left(\frac{\delta g_{MN}}{\delta X^K}
\partial_\alpha X^M\partial_\beta X^N+
\frac{\delta b_{MN}}{\delta X^K}
\partial_\alpha X^M\partial_\beta X^N\right)
\mati^{\beta\alpha}\sqrt{\det\mat}+
\nonumber \\
+T_{S(p-1)}\partial_\alpha\left[e^{-\Phi}
g_{KM}\partial_\beta X^M
\mati_S^{\beta\alpha}
\sqrt{\det\mat}\right]
+\nonumber \\
+T_{S(p-1)}\partial_\alpha \left[
e^{-\Phi}b_{KM}\partial_\beta X^M
\mati^{\beta\alpha}_A
\sqrt{\det\mat}\right]+\tilde{J}_K=0  \ ,
\nonumber \\
\end{eqnarray}
where
\begin{eqnarray}\label{currentxkbps}
\tilde{J}_K &=& \frac{\delta S_{WZ}} {\delta X^K}\nonumber \\
&&=\mu_{S(p-1)}\sum_{n\geq 0}
\frac{1}{n!(2!)^n(2p-2n)!} \epsilon^{\alpha_1\dots \alpha_p}
\left[\partial_Kb_{MN}\partial_{\alpha_1}X^M
\partial_{\alpha_2}X^N(\mtF)^{n-1}_{\alpha_3
\dots \alpha_{2n}}
\mtC_{\alpha_{2n+1}\dots \alpha_p}\right.
\nonumber \\
&&\left.+(\mtF)^n_{\alpha_1\dots \alpha_{2n}}
\partial_K \mtC_{M_1\dots M_{2p-2n}}
\partial_{\alpha_{2n+1}}X^{M_1}\dots
\partial_{\alpha_p}X^{M_{2p-2n}}-\right.
\nonumber \\
&&\left.-2n\partial_{\alpha_1}
\left[b_{KM}\partial_{\alpha_2}
X^M(\mtF)^{n-1}_{\alpha_3\dots
\alpha_{2n}}
\mtC_{\alpha_{2n+1}\dots \alpha_p}
\right]-\right. \nonumber \\
&&\left.-(2p-2n)\partial_{\alpha_{2n+1}}
\left[(\mtF)^{n}_{\alpha_1\dots \alpha_{2n}}
C_{KM_2\dots M_{2p-2n}}\partial_{\alpha_{2n+2}}
X^{M_2}\dots \partial_{\alpha_p}X^{M_{2p-2n}}
\right]\right]\ .
\nonumber \\
\end{eqnarray}
In the same way we get
that the equation of
 motion for $A_\alpha$ are
\begin{equation}\label{eqAbps}
T_{S(p-1)}\partial_\beta
\left[e^{-\Phi}
\mati^{\alpha\beta}_A
\sqrt{-\det\mat}\right]+\tilde{J}^\alpha=0 \ ,
\end{equation}
where
\begin{equation}\label{currentAbps}
\tilde{J}^{\alpha_1}=
\mu_{S(p-1)}\sum_{n\geq 0}
\frac{2n}{n! 2^n (2p-2n)!}
\epsilon^{\alpha_1\dots\alpha_p}
\partial_{\alpha_2}\left[ (\mtF)^{n-1}_{\alpha_3
\dots \alpha_{2n}}
\mtC_{\alpha_{2n+2}\dots \alpha_p}
\right] \  .
\end{equation}
In what follows we will
proceed in the  same  way as in
\cite{Kluson:2005fj} so we
can be brief and recommend
the paper
\cite{Kluson:2005fj} for more details.

We begin with the presumption that the tachyon kink depends on the
time coordinate on the worldvolume of Dp-brane. We will also see
that when we consider the singular limit we obtain the formal
solution that leads to the negative expression under square root.
In spite this fact we will argue that this singular solution
describes S(p-1)-brane.

More precisely, let us consider the following ansatz for tachyon
\cite{Sen:2003tm}
\begin{equation}\label{ansT}
T(x,\xi) = f(a(x-t(\xi)) \ ,
\end{equation}
where $x$ is time coordinate on the worldvolume of Dp-brane and
where $t$ is some unknown function of the $\xi^\alpha \ ,
\alpha=1,\dots,p$ euclidean coordinates on the kink. We also
presume as in \cite{Sen:2003tm} that $f(u)$ satisfies following
properties
\begin{equation}
f(-u)=-f(u) \ , f'(u)>0 \ , \forall u \ ,
f(\pm \infty)=\pm \infty
\
\end{equation}
but is otherwise an arbitrary function of its argument $u$. $a$ is
a constant that we shall take to $\infty$ in the end. In this
limit we have $T=\infty$ for $x>t(\xi)$ and $T=-\infty$ for
$x<t(\xi)$. Let us also presume following ansatz for the massless
fields
\begin{equation}\label{ansA}
X^M(x,\xi)=X^M(\xi) \ ,
A_x(x,\xi)=0 \ , A_{\alpha}(x,\xi)=
A_\alpha(\xi) \ ,
\alpha=1,\dots,p \ .
\end{equation}
With this ansatz the
matrix $\bA_{\mu\nu}$ takes the form
\begin{equation}
\bA=\left(\begin{array}{cc}
a^2f'^2 & -a^2f'^2\partial_\beta t \\
-a^2f'^2\partial_\alpha t &
\mat_{\alpha\beta}+
a^2f'^2\partial_\alpha t
\partial_\beta t
\end{array}\right) \ ,
\end{equation}
where
\begin{equation}
\mat_{\alpha\beta}=
(g_{MN}+b_{MN})
\partial_\alpha X^M\partial_\beta X^N+
F_{\alpha\beta} \ .
\end{equation}
Now using the fact that
\begin{equation}
\det\bA=
\det(\bA_{\alpha\beta}-
\bA_{\alpha x}\frac{1}{\bA_{xx}}
\bA_{x\beta})\det\bA_{xx}
\end{equation}
we get
\begin{equation}
\det\bA=a^2f'^2\det\mat \ .
\end{equation}
As a next step we  determine
the inverse matrix $\bAi$. After some
calculations we get the result
\begin{eqnarray}\label{bai}
\bAi^{\alpha\beta}=\mati^{\alpha\beta} \
\bAi^{x\beta}=\partial_\alpha t
\mati^{\alpha\beta} \ , \nonumber \\
\bAi^{\alpha x}=\mati^{\alpha \beta}
\partial_\beta t \ ,
\bAi^{xx}=\partial_\alpha t\mati^{\alpha\beta}
\partial_\beta t \  \nonumber \\
\end{eqnarray}
For next purposes following relation
will be also  useful
\begin{equation}\label{baip}
\bAi^{\mu x}_S-\bAi^{\mu\alpha}_S
\partial_\alpha t=\frac{1}{a^2f'^2}
\left(\delta^{\mu}_x-\bAi^{x\mu}_S\right) \ .
\end{equation}

With the help of this expression we get
\begin{eqnarray}
\partial_\mu\left[e^{-\Phi}V
\partial_\nu T\bAi^{\nu\mu}_S
\sqrt{-\det\bA}\right]=
 \nonumber \\
=V'af'e^{-\Phi}\sqrt{-\det\mat}
-V\partial_\alpha
\left[e^{-\Phi}\mati^{\beta\alpha}_S
\partial_\beta t\sqrt{-\det\mat}\right] \ ,
\nonumber \\
\end{eqnarray}
where we have used the
fact that the only field that depends on
$x$ is a tachyon. Then  the DBI part
\footnote{"DBI" part of the equation
of motion means the part that arises
from the variation of the DBI action.}
 of the
 tachyon equation of motion (\ref{eqT})
takes the form
\begin{equation}
\partial_\alpha\left[
V(f)e^{-\Phi}\mati^{\beta\alpha}_S
\partial_\beta t\sqrt{-\det\mat}\right]=i
\partial_\alpha\left[
V(f)e^{-\Phi}\mati^{\beta\alpha}_S
\partial_\beta t\sqrt{\det\mat}\right] \ .
\end{equation}
Now we consider the DBI part of the equation of motion for $X^K$
(\ref{eqX}). In the same way as in \cite{Kluson:2005fj}
we  can show the the DBI part
of the equation of motion
(\ref{eqX})  takes the form
\begin{eqnarray}\label{dbixf}
\sqrt{-1}af'V
\left(-\partial_K[e^{-\Phi}]
\sqrt{\det\mat}-\frac{e^{-\Phi}}{2}
\left(g_{MN,K}+b_{MN,K}\right)
\partial_\alpha X^M\partial_\beta X^N
\mati^{\alpha\beta}
\sqrt{\det\mat}\right.
\nonumber \\
\left.
+\partial_\beta\left[
e^{-\Phi}g_{KM}\partial_\alpha
X^M \mati^{\alpha\beta}_S
\sqrt{\det\mat}\right]+
\partial_\beta\left[e^{-\Phi}
b_{KM}\partial_\alpha X^M\mati^{\beta\alpha}_A
\sqrt{\det\mat}\right]\right) \ .
\nonumber \\
\end{eqnarray}
Now let us consider the DBI part
of the
 equation of motion for gauge field
(\ref{eqA}). For
$A_x$ we get
\begin{eqnarray}
\partial_\nu\left[Ve^{-\Phi}\bAi
^{x\nu}_A\sqrt{-\det\bA}\right]
=\sqrt{-1}af'V \partial_\alpha
\left[e^{-\Phi}\mati^{\beta\alpha}_A\partial_\beta t
\sqrt{\det\mat}\right] \ .
\nonumber \\
\end{eqnarray}
On the other hand the equations of motion for $A_\alpha$
take the form
\begin{eqnarray}\label{dbiaf}
\partial_\mu\left[e^{-\Phi}
\bAi^{\alpha \mu}_A\sqrt{-\det\bAi}\right]=
\sqrt{-1}af'V\partial_\beta\left[
e^{-\Phi}\mati^{\alpha\beta}_A\sqrt{\det\mat}\right] \ .
\nonumber \\
\end{eqnarray}



As a next step we evaluate the
currents $J_T,J_K$ and $J^{\mu_1}$.
 for the
ansatz (\ref{ansT}) and (\ref{ansA}).
It was shown in \cite{Kluson:2005fj}
 following
\cite{Skenderis:2002vf}
that these currents take the form
\begin{equation}\label{currentAh}
J^{\mu_1}=
\sum_{n\geq 0}
\frac{2n}{n! 2^n(2p-2n)!}
\epsilon^{\mu_1\dots \mu_{p+1}}
\partial_{\mu_2}\left[V(T) (\mF)^{n-1}_{\mu_3\dots \mu_{2n}}
C_{\mu_{2n+1}\dots \mu_{p}}\partial_{\mu_{p+1}}T
\right] \ ,
\end{equation}
\begin{eqnarray}\label{currentTh}
J_T=\sum_{n\leq 0}
\frac{1}{n!(2!)^n (2p-2n)!}
\epsilon^{\mu_1\dots \mu_{p+1}}
V'(T)\left((\mF)^n_{\mu_1\dots \mu_{2n}}
C_{\mu_{2n+1}\dots \mu_p}
\partial_{\mu_{p+1}}T\right)-\nonumber \\
-\partial_{\mu_{p+1}}
\sum_{n\leq 0}
\frac{1}{n!(2!)^n (2p-2n)!}
\epsilon^{\mu_1\dots \mu_{p+1}}
\left[V(T)(\mF)^n_{\mu_1\dots \mu_{2n}}
C_{\mu_{2n+1}\dots \mu_p}\right] \
\nonumber \\
\end{eqnarray}
and
\begin{eqnarray}\label{currentXKh}
J_K&=& \sum_{n\leq 0} \frac{1}{n!(2!)^n (2p-2n)!} \epsilon^{\mu_1\dots
\mu_{p+1}} \left[V(T)b_{MN,K}\partial_{\mu_1}X^M
\partial_{\mu_2}X^N(\mF)^{n-1}_{\mu_3
\dots \mu_{2n}}
C_{\mu_{2n+1}\dots \mu_p}\partial_{\mu_{p+1}}
T\right.
\nonumber \\
&&\left.+V(T)(\mF)^n_{\mu_1\dots \mu_{2n}}
\partial_K C_{M_1\dots M_{2p-2n}}
\partial_{\mu_{2n+1}}X^{M_1}\dots
\partial_{\mu_p}X^{M_{2p-2n}}\partial_{\mu_{p+1}}
T-\right.
\nonumber \\
&&\left.-2\partial_{\mu_1} \left[V(T)b_{KM}\partial_{\mu_2}
X^M(\mF)^{n-1}_{\mu_3\dots \mu_{2n}} C_{\mu_{2n+1}\dots
\mu_p}\partial_{\mu_{p+1}}T
\right]-\right. \nonumber \\
&&\left.-(2p-2n)\partial_{2n+1} \left[V(T)(\mF)^{n}_{\mu_1\dots
\mu_{2n}} C_{KM_2\dots M_{2p-2n}}\partial_{\mu_{2n+2}} X^{M_2}\dots
\partial_{\mu_p}X^{M_{2p-2n}}
\partial_{\mu_{p+1}}T\right]\right] \ .
\nonumber \\
\end{eqnarray}
Let us start with $J_T$ that can be written as
\begin{eqnarray}
J_T=
-\sum_{n\leq 0}V(T)\frac{1}{n!(2!)^n (2p-2n)!}
\epsilon^{\mu_1\dots \mu_{p+1}}
\partial_{\mu_{p+1}}\left((\mF)^n_{\mu_1\dots \mu_{2n}}
C_{\mu_{2n+1}\dots \mu_p}
\right) \ .
\nonumber \\
\end{eqnarray}
As the first step we determine the components
of the embedding of various fields. It can
be shown \cite{Kluson:2005fj} that
the only
nonzero components of $\mF_{\mu\nu}$ are
$\mF_{\alpha\beta}$.
For $C^{(n)}$ the situation
is the same, namely any component with $x$
index is equal to zero.
Then it can be easily shown
that the  tachyon
current is equal to zero \cite{Kluson:2005fj}.

Now we consider the gauge current $J^{\mu}$.
Firstly, it can be easily
shown \cite{Kluson:2005fj}
that $J^x$ is equal to
\begin{eqnarray}
J^x
=-af'V\sum_{n\geq 0}
\frac{2n}{n! 2^n (2p-2n)!}
\epsilon^{x\alpha_1\dots \alpha_p}
\partial_{\alpha_1}\left[(\mF)^{n-1}_{\alpha_3
\dots \alpha_{2n}}
C_{\alpha_{2n+1}\dots \alpha_{p-1}}\partial_{\alpha_p}t\right]
\nonumber \\
\end{eqnarray}
while $J^{\alpha_1}$
takes the form
\begin{eqnarray}\label{currentAf}
J^{\alpha_1}
=\sum_{n\geq 0} af'V\frac{2n}{n! 2^n (2p-2n)!}
\epsilon^{\alpha_1\alpha_2\dots \alpha_p x}
\partial_{\alpha_2}\left[(\mF)^{n-1}_{\mu_3\dots \mu_{2n}}
C_{\mu_{2n+1}\dots \mu_{p}}\right]=
af'V\tilde{J}^{\alpha_1} \ ,  \nonumber \\
\end{eqnarray}
where
we have
introduced the notation  $\tilde{J}^{\alpha_1}$
that is a correct
form of the gauge current for S(p-1)-brane.
If we now combine
(\ref{dbiaf}) with (\ref{currentAf}) we get
\begin{equation}\label{eqaf}
af'V
\left[i\partial_\beta\left[
e^{-\Phi}\mati^{\alpha\beta}_A\sqrt{\det\mat}\right]+
\tilde{J}^{\alpha}\right]=0 \ .
\end{equation}
Let us now analyze
the behavior of the term  $af'V$ in the limit
$a\rightarrow \infty$.
Since by definition $f'(u)$ is finite for
all $u$ it remains
to study the properties of the expression $aV$.
Since $V\sim e^{-T}$ for $T\rightarrow \infty$ we have
\begin{eqnarray}
\lim_{a\rightarrow \infty} aV(f(a(x-t(\xi))=
 \ (\mathrm{for} \ x\neq t(\xi))
 \nonumber \\
 \lim_{a\rightarrow \infty}
 \frac{a}{e^{f(a(x-t(\xi)))}}=
\frac{1}{(x-t(\xi))f'}\lim_{a\rightarrow \infty}
e^{-f(a(x-t(\xi)))}=0 \nonumber \\
\end{eqnarray}
We see that for $x\neq t(\xi)$
the expression $aV$ goes to zero in
the limit $a\rightarrow \infty$.
On the other hand for $x=t(\xi)$
the potential $V(0)=\tau_p$ and hence in order to obey the
equation of motion for $A_{\alpha}$
we find that the expression in
the bracket in (\ref{eqaf}) should vanish.
In fact, this
expression is correct form of the equation of motion for
$A_{\alpha}$ that propagate on
the worldvolume of S(p-1)-brane.

From (\ref{eqaf}) we
can also deduce that  the S(p-1)-brane
tension and its charge
with respect to Ramond-Ramond
fields are equal to
\begin{equation}\label{Ts}
T_{S(p-1)}=iT_{p-1}
\ ,
\mu_{S(p-1)}=\mu_{p-1} \ ,
\end{equation}
where $T_{p-1}$ is tension
of BPS D(p-1)-brane and
$\mu_{p-1}$ is its corresponding
charge. Even if the form of
the equation (\ref{eqaf})
 suggests
that the tension of S(p-1)-brane
can be arbitrary we will give
arguments for the validity
of (\ref{Ts}) in the next
subsection.

Let us now turn to the equation
of motion for $A_x$. It was
shown in \cite{Kluson:2005fj}
that it has solution in case
when we demand that
\begin{equation}
\partial_\alpha t=0 \ .
\end{equation}
The expression above also
implies that the equation
of motion for tachyon that
for $J_T=0$ has the form
\begin{equation}
V\partial_\alpha\left[\sqrt{-1}e^{-\Phi}
\mati^{\beta\alpha}_S\partial_\beta t\sqrt{\det\mat}
\right]=0 \
\end{equation}
 is obeyed.


Finally, we will consider the current
$J^K$. Again, as was shown in
\cite{Kluson:2005fj} that
for the ansatz (\ref{ansT}) and (\ref{ansA})
this current takes the form
\begin{eqnarray}\label{currentXKf}
J_K=af'V\sum_{n\leq 0}
\frac{1}{n!(2!)^n (2p-2n)!}
\epsilon^{\alpha_1\dots \alpha_px}
\left(b_{MN,K}\partial_{\alpha_1}X^M
\partial_{\alpha_2}X^N(\mF)^{n-1}_{\alpha_3
\dots \alpha_{2n}}
C_{\alpha_{2n+1}\dots \alpha_p}
\right.
\nonumber \\
\left.+
(\mF)^n_{\alpha_1\dots \alpha_{2n}}
\partial_K C_{M_1\dots M_{2p-2n}}
\partial_{\alpha_{2n+1}}X^{M_1}\dots
\partial_{\alpha_p}X^{M_{2p-2n}}-\right.
\nonumber \\
\left.-2\partial_{\alpha_1}
\left[b_{KM}\partial_{\alpha_2}
X^M(\mF)^{n-1}_{\alpha_3\dots
\alpha_{2n}}
C_{\alpha_{2n+1}\dots \alpha_p}
\right]+\right.
 \nonumber \\
\left.+ (2p-2n) \partial_{\alpha_{2n+1}}
\left[(\mF)^{n}_{\alpha_1\dots \alpha_{2n}}
C_{KM_2\dots M_{2p-2n}}\partial_{\alpha_{2n+2}}
X^{M_2}\dots \partial_{\alpha_p}X^{M_{2p-2n}}
\right]\right)\equiv af'V\tilde{J}^K \ .  \nonumber \\
\end{eqnarray}
Using (\ref{dbixf})
and (\ref{currentXKf}) we obtain the final
form of the equation of motion for $X^K$ in  the form
\begin{eqnarray}\label{eqXKf}
&& af'V \left(-i\partial_K[e^{-\Phi}]
\sqrt{\det\mat}-i\frac{e^{-\Phi}}{2}
\left(g_{MN,K}+b_{MN,K}\right)
\partial_\alpha X^M\partial_\beta X^N
\mati^{\alpha\beta}
\sqrt{\det\mat}\right. \nonumber \\
&&\left. +i\partial_\beta\left[
e^{-\Phi}\left\{g_{KM}\partial_\alpha X^M \mati^{\alpha\beta}_S +
b_{KM}\partial_\alpha X^M\mati^{\beta\alpha}_A\right\}
\sqrt{\det\mat}\right] +\tilde{J}^K\right)=0 \ .
\nonumber \\
\end{eqnarray}
Following now a
discussion given below (\ref{eqaf}) we see that
the expression in
the bracket in (\ref{eqXKf}) should be equal to
zero. On the other hand
this equation is exactly the equation of
motion for the embedding mode
that lives on the worldvolume of
S(p-1)-brane.

Let us briefly discuss the meaning of the condition
$\partial_\alpha t=0$. Following \cite{Kluson:2005fj} we can argue
that all tachyon kink solutions are parameterized with the
constant $t$ that determines the core of the kink and that all $t$
are equivalent. This is natural result since we have not fixed the
gauge on the worldvolume of non-BPS Dp-brane.


\subsection{Stress energy tensor}
Further support for an interpretation
of the tachyon kink as a
lower dimensional S(p-1)-brane
can be derived from the analysis of
the stress energy tensor for the non-BPS Dp-brane.
In order to
find its form  recall that we can
write the action (\ref{acg}) as
\begin{equation}\label{dactem}
S_{p}=-\int d^{10}xd^{(p+1)}
\xi\delta
(X^M(\xi)-x^M)e^{-\Phi}V(T)
\sqrt{-\det \bA} \ .
\end{equation}
From  (\ref{dactem}) we can
easily determine components of the
stress energy tensor $T_{MN}(x)$ of an
unstable D-brane using the
fact that the stress energy
tensor $T_{MN}(x)$  is defined as the
variation of $S_p$ with respect to $g_{MN}(x)$
\begin{eqnarray}\label{TMNg}
T_{MN}(x)
=-2
\frac{\delta S_{p}}{
\sqrt{-g(x)}\delta g^{MN}(x)}=\nonumber \\
=-\int d^{(p+1)}\xi\frac{\delta(X^M(\xi)
-x^M)}{\sqrt{-g(x)}}e^{-\Phi}V g_{MK}g_{NL}
\partial_{\mu}X^K\partial_{\nu}
X^L(\bA^{-1})^{\nu\mu}_S
\sqrt{-\det \bA} \ . \nonumber \\
\end{eqnarray}
Now from (\ref{ansT}) and (\ref{ansA})
we  know that all massless modes are $x$ independent.
Hence (\ref{TMNg}) is equal to
\begin{eqnarray}
&&T_{MN}(x)=-\int
dx af'V(f(x)) \int d^{p}\xi\frac{\delta(X^M(\xi)
-x^M)}
{\sqrt{-g(x)}}\times \nonumber \\
&&\times e^{-\Phi} g_{MK}g_{NL}
\partial_{\alpha}X^K\partial_{\beta}
X^L\mati^{\beta\alpha}_S
\sqrt{-\det \mat}= \nonumber \\
&&-iT_{p-1} \int d^{p}\xi\frac{\delta(X^M(\xi) -x^M)}
{\sqrt{-g(x)}}e^{-\Phi} g_{MK}g_{NL}
\partial_{\alpha}X^K\partial_{\beta}
X^L\mati^{\beta\alpha}_S
\sqrt{\det \mat} \ ,    \nonumber \\
\end{eqnarray}
where
\begin{equation}
T_{p-1}=\int dx aV(f)f'=
\int dm V(m)
\end{equation}
is a tension of BPS D(p-1)-brane.
In other words the stress energy
tensor evaluated on the ansatz (\ref{ansT})
and (\ref{ansA})
corresponds to the stress
energy tensor for S(p-1)-brane.
We also see that it is natural
to define the tension of $S(p-1)$-brane
as $T_{S(p-1)}=iT_{p-1}$.

In the same way we can calculate the
charge to NS-NS two form fields. Again,
since this charge follows from the
variation of the DBI part of the
non-BPS Dp-brane effective action
we again get that this charge is
imaginary. In conclusion, the effective
field theory analysis of the time
dependent tachyon kink suggests that
S(p-1)-brane has imaginary charge
with respect to NS-NS fields while
is real to Ramond-Ramond fields.

\section{S0-brane in the T-dual
background}
Now we would like to apply the
general discussion given in previous
section  for
S0-brane in the dual background defined
in section (\ref{third}).
 Recall that S0-brane action
has the form
\begin{equation}\label{Ss0}
S=-\tau_{_{S0}}\int
d\xi e^{-\Phi}\sqrt{g_{MN}\dot{X}^M\dot{X^N}} \ ,
\end{equation}
where $\xi$ is world line coordinate and
$X^M$ are embedding
coordinates of S0-brane.
If we vary the action
(\ref{Ss0}) we obtain
the equations of motion for
$X^K$ in the form
\begin{equation}
\partial_K[e^{-\Phi}]
\sqrt{g_{MN}\dot{X}^M\dot{X^N}}
+\frac{e^{-\Phi}\partial_Kg_{MN}
\dot{X}^M\dot{X^N}}
{2\sqrt{g_{MN}\dot{X}^M\dot{X^N}}}
-\frac{d}{d\xi}
\left[\frac{e^{-\Phi}g_{MN}\dot{X}^N}
{\sqrt{g_{MN}\dot{X}^M\dot{X^N}}}
\right]=0 \ .
\end{equation}
Since we presume that
S0-brane wraps the $z$ direction
we choose
the gauge
\begin{equation}
\xi=Z \
\end{equation}
and hence
\begin{equation}
g_{MN}\dot{X}^M\dot{X}^N=
g_{zz}+2g_{tz}\dot{T}+
g_{tt}\dot{T}^2+
g_{\rho\rho}\dot{\rho}^2 \ ,
\end{equation}
where $X^0\equiv T \ , \dot{(\dots)}= \frac{d(\dots)}{d\xi}$. For
convenience we write again the background fields in T-dual
spacetime
\begin{eqnarray}
g_{zz}=\frac{\alpha'^2}{L^2\sinh^2\rho} \ , g_{tt}=-L^2 \ ,
g_{tz}=g_{zt}=\alpha' \ ,
g_{\rho\rho}=L^2 \ ,\nonumber \\
b_{zt}=-L^2 \ ,
e^{-\Phi}=\frac{L}{g_s\sqrt{\alpha'}}
\sinh\rho \ . \nonumber \\
\end{eqnarray}
Then the action (\ref{Ss0}) takes the form
\begin{equation}\label{S0}
S=-\tau_{S0}\int dz
e^{-\Phi}\sqrt{g_{zz}+2g_{tz}\dot{T}+
g_{tt}\dot{T}^2+
g_{\rho\rho}\dot{\rho}^2} \ .
\end{equation}
Let us now try to
find the `dynamics' of S0-brane in $T$-dual
background. Note that the Lagrangian has the form
\begin{equation}
\mL=-\sqrt{V+\sum_i (f_i(\partial_0{\Phi}^i)^2
+B_i\partial_0\Phi^i)}\equiv -\triangle \ ,
\end{equation}
where $V$ contain scalar potential
for various fields $\Phi^i$.
Then in the same way as in the
section (\ref{third})
we determine the corresponding
Hamiltonian
\begin{equation}\label{HS0}
H=\sqrt{\left(V-\sum_i\frac{B_i^2}{4f_i}\right)
\left(1-\sum_i\frac{P^2_i}
{f_i}\right)} -\sum_i\frac{B_iP_i}{2f_i} \ ,
\end{equation}
where $P_i$ is momentum
conjugate to $\Phi^i$.
Now from (\ref{S0}) we have
\begin{equation}
V=\tau_{S0}^2e^{-2\Phi}g_{zz} \ ,
f_T=\tau_{S0}^2e^{-2\Phi}g_{tt} \ ,
b_T=2\tau_{S0}^2e^{-2\Phi}g_{tz} \ ,
f_\rho=\tau_{S0}^2e^{-2\Phi}
g_{\rho\rho} \
\end{equation}
and hence the Hamiltonian takes
the form
\begin{equation}
H=\sqrt{
\left(\frac{g_{zz}g_{tt}-g_{tz}^2}
{g_{tt}}\right)\left(-\frac{P^2_T}
{g_{tt}}-\frac{P^2_\rho}{g_{\rho\rho}}
+e^{-2\Phi}\tau_{S0}^2\right)}
-\frac{g_{tz}}{g_{tt}}P_T \ .
\end{equation}
As usual the equation of motion
for $T$ that follows from
(\ref{HS0})
 implies that $P_T=const$.
On the other hand
the equation of
motion for $\rho$ is equal
to
\begin{equation}\label{dotrhos}
\frac{d\rho}{d\xi}=\frac{\delta H}{\delta P_\rho}=
\left(\frac{g_{tz}^2-g_{zz}g_{tt}}
{g_{tt}}\right)\frac{P_\rho}
{g_{\rho\rho}\sqrt(\dots)} \ .
\end{equation}
If we again express $P_\rho$ from
the Hamiltonian (\ref{HS0}) and
use the fact that it is
conserved (In a sense
that $\frac{dH}{dz}=0$) we obtain
\begin{eqnarray}
P^2_\rho=
-\frac{L^4\sinh^2\rho}{\alpha'^2\cosh^2\rho}
(E-\frac{\alpha'}{L^2}
P_T)^2+P_T^2+\frac{L^4\tau_{S0}^2}{\alpha'g_s^2}
\sinh^2\rho \ .
\nonumber \\
\end{eqnarray}
Using this expression
the equation (\ref{dotrhos})
is equal to
\begin{eqnarray}\label{dotrhos0a}
\left(\frac{d\rho}{d\xi}
\right)^2=
\left(\frac{g_{tz}^2-g_{zz}g_{tt}}
{g_{tt}}\right)^2\frac{P^2_\rho}
{g_{\rho\rho}^2
(E+\frac{g_{zt}}{g_{tt}}P_T)^2}=
\nonumber \\
-\frac{\alpha'^2}
{L^4}
\frac{\cosh^2\rho}{\sinh^2\rho}
+\frac{\alpha'^4}{L^8}\frac{\cosh^4\rho}
{\sinh^4\rho
(E-\frac{\alpha'}{L^2}P_T)^2}
\left(P_T^2+\frac{L^4\tau_{S0}^2}
{\alpha'g_s^2}
\sinh^2\rho
\right)\nonumber \\
\end{eqnarray}
using
\begin{eqnarray}
\frac{g_{tz}^2-g_{zz}g_{tt}}
{g_{tt}g_{\rho\rho}}=-\frac{\alpha'^2}
{L^4}\frac{
\cosh^2\rho}{\sinh^2\rho} \ . \nonumber \\
\end{eqnarray}
To compare the ``dynamics'' of
S0-brane with the dynamics of
D0-brane given in equation (\ref{dotrhod0})
we have to take into account  the
parametrization of $\rho$.
Namely, in (\ref{dotrhod0})
 $\rho$ is the function of $t$
that was identified with
$X^0$ while in the equation
(\ref{dotrhos0a})
 $\rho$ is the function of $\xi$ that
is identified with $z$.
On the other hand we
can certainly write
\begin{equation}\label{rhots}
\frac{d\rho}{d\xi}= \frac{d\rho}{dT}\frac{dT}{d\xi}
=\frac{d\rho}{dT}\frac{\delta H}{\delta P_T} \ ,
\end{equation}
where
\begin{eqnarray}\label{hpt}
\frac{dH}{dP_T}=
\frac{\alpha'}{L^2}\frac{\frac{\alpha'}
{L^2}P_T+E\sinh^2\rho}
{\sinh^2\rho(E-\frac{\alpha'}{L^2}P_T)}
\ .  \nonumber \\
\end{eqnarray}
If we combine (\ref{dotrhos0a}),
(\ref{rhots}) and (\ref{hpt})
 we get
\begin{eqnarray}\label{dotrhoT}
\left(\frac{d\rho}{dT}\right)^2=
\cosh^2\rho-
\cosh^4\rho\sinh^2\rho
\frac{
\left[\frac{L^4}{\alpha'^2}
E^2-\frac{L^4\tau_{S0}^2}{\alpha'g_s^2}\right]}
{(P_T+\frac{L^2}{\alpha'}E\sinh^2\rho)^2} \ .
\nonumber \\
\end{eqnarray}
However one can check that this is exactly the same
differential equation for $\rho$  as in
case of D0-brane given
in (\ref{dotrhod0}) when we note that $\tau_{_{S0}}=i\tau_0$. Let us
now try to insert the solution $\cosh\rho\cos T=C$ into the
equation (\ref{dotrhoT}). After some calculations we get
\begin{eqnarray}
1-\frac{1}{C^2}=
\frac{\sinh^4\rho \left[\frac{L^4}{\alpha'^2}
E^2-\frac{L^4\tau_{S0}^2}{\alpha'g_s^2}\right]}
{(P_T+\frac{L^2}{\alpha'}E\sinh^2\rho)^2} \ .
\nonumber \\
\end{eqnarray}
Since the left side is a constant it is clear that the only way
how this equation is obeyed is to demand that $P_T=0$. This is in
complete agreement with the previous sections since $P_T$ is
canonical conjugate to $X^0$.
Then the equation above
implies
\begin{equation}\label{Es}
E^2=C^2\frac{\alpha'\tau_{_{S0}}^2}{g_s^2} \ .
\end{equation}
From the point of view
of S0-brane worldvolume theory this is
perfectly consistent
result since now $E$ is real. However from
the point of view of original
theory where $\tau_{_{S0}}=i\tau_0$
we obtain imaginary $E$
which of course is expected since
$\tau_{S0}$ is imaginary.

It is  also interesting to study the
dependence of $T$ on $\xi$. In fact, using
(\ref{HS0}) we obtain
\begin{equation}
\frac{dT}{d\xi}=
\frac{\delta H}{\delta P_T}=
\frac{g_{zt}^2-g_{zz}g_{tt}}{g_{tt}^2}
\frac{P_T}{E+\frac{g_{tz}}{g_{tt}}}-
\frac{g_{tz}}{g_{tt}}
\end{equation}
that for  $P_T=0$ implies
\begin{equation}
\frac{dT}{d\xi}=-\frac{g_{tz}}{g_{tt}}=
\frac{\alpha'}{L^2} \ .   
\end{equation}
We see that the S0-brane does
not take the fixed position in time, rather
the dependence of $T$ on $\xi=z$ is in 
perfect agreement with the
result (\ref{dotZE0}).

On the other hand, if we insert
(\ref{Es}) into (\ref{dotrhos0a})
and use $P_T=0$ we obtain a
differential equation for $\rho$
in the form
\begin{equation}
\left(\frac{d\rho}{d\xi}\right)^2=
\frac{\alpha'^2}{L^4}
\left(\frac{\cosh^4\rho}{C^2\sinh^2\rho}-
\frac{\cosh^2\rho}{\sinh^2\rho}\right)
 \ .
\end{equation}
Even if this equation can
be solved explicitly we restrict
ourselves to the case when
$C=1$ in order  to demonstrate the main
properties of given solution.
For $C=1$ the equation above
reduces into
\begin{equation}
\frac{d\rho}{d\xi}=
\frac{\alpha'}{L^2}\cosh\rho \
\end{equation}
that has the solution
\begin{equation}\label{rhod}
\sinh\rho=\left|\tan\left(\frac{\alpha'}{L^2}
\xi\right)\right| \ ,
\end{equation}
where we have chosen the integration constant in such a way that
for $ \xi=0$, $\rho=0$. The solution (\ref{rhod}) describes
S0-brane that many times wraps $\xi=z$ direction before it reaches
its maximum value at  $\xi_{max} =\frac{L^2\pi}{2\alpha'}$ (Note
that $L^2\gg \alpha'$ in order to trust supergravity solution.).
Then it again spirals down until it reaches the point $\rho=0$ at
$\xi_f=\frac{L^2}{\alpha'}\pi$.

\section{Summary and Conclusion}
In this paper, we have studied the unphysical dS$_2$-branes in the
covering space of the SL(2,R) WZW model, that is the AdS$_3$ space
time, supported by NS-NS three-form flux, and observed in the
$T$-dual set up, the emergence of the S-branes. We have been able
to present a physical interpretation of the unphysical solutions
with imaginary electric flux corresponding to dS$_2$ branes. This
becomes clear in the T-dual picture, in the form of an S0-brane
that arises from the time dependent tachyon condensation on an
unstable D1-brane. We have also been able to show that the
previously found unphysical solutions correspond, in fact, to
perfect and acceptable solutions in string theory (even if the
initial configurations of tachyon that corresponds to S-brane have
to be fine tuned) since they arise from the singular, time dependent
tachyon condensation.
We have also shown that these S-branes couple
to imaginary NS-NS fields, but to real R-R fields, and hence in
the terminology of \cite{Durin:2005ts}, correspond to S$^{-}$
branes. We further have analyzed the time dependent tachyon
condensation on non-BPS Dp-branes in general background and found
out a class of time dependent singular solutions which correspond
to the S$^-$ branes. Arguments in favor of this have also been
given by studying the stress tensor, that revealed the fact that
indeed these S-branes couple to imaginary NS-NS fields, but to
real R-R fields. There are further directions of research that one
can adopt. One of them is to analyze the unphysical branes
\cite{Hikida:2005vd} in the Nappi-Witten model. The DBI action on
the unphysical branes in that background have been shown to be
imaginary, and hence in the present context, might correspond to
some kind of Dirichlet S-branes. One can further analyze the one
loop partition function for the branes in the above backgrounds,
and give interpretations in the same spirit of
\cite{Durin:2005ts}. We hope to come back to some of these issues
in near future.

\vskip .5cm
\noindent{\bf Acknowledgements:}
\vskip .2cm
\noindent
The work of JK was supported  in part by the Czech Ministry of
Education under Contract No. MSM 0021622409, by INFN, by the
MIUR-COFIN contract 2003-023852, by the EU contracts
MRTN-CT-2004-503369 and MRTN-CT-2004-512194, by the INTAS contract
03-516346 and by the NATO grant PST.CLG.978785.The work of RRN was
supported by INFN. The work KLP was supported partially by PRIN
2004 - "Studi perturbativi e non perturbativi in teorie
quantistiche dei campi per le interazioni fondamentali".


\end{document}